\documentclass[conference, pdftex]{IEEEtran}
\usepackage{cite}
\usepackage{graphicx}
\usepackage{colortbl}
\usepackage[table]{xcolor}
\usepackage{xurl}
\usepackage{fontawesome5}
\usepackage{hyperref}
\usepackage{microtype}

\usepackage{cleveref}
\crefname{figure}{Fig.}{Figs.}
\Crefname{figure}{Figure}{Figures}
\crefname{table}{Table}{Tables}
\crefname{section}{Section}{Sections}
    
\newcommand{\Circled}[1]{\textcircled{{\footnotesize #1}}}
\newcommand{\sourcecode}[1]{\texttt{#1}}
\newcommand{\refactoring}[1]{\textit{#1}}
\newcommand{\microchange}[1]{\textit{#1}}

\definecolor{classRefactoringColor}{RGB}{129, 199, 212}
\definecolor{methodRefactoringColor}{RGB}{51, 166, 184}
\definecolor{modificationColor}{RGB}{247, 225, 59}
\definecolor{additionColor}{RGB}{202,244,202}
\definecolor{removalColor}{RGB}{255, 0, 0}
\definecolor{variableRefactoringColor}{RGB}{13, 86, 97}
\definecolor{microChangeColor}{RGB}{128, 0, 128}

\newcommand{\ColorBox}[1]{\raisebox{0.2em}{\fcolorbox{black}{#1}{~~~}}}
\newcommand{\ShowColor}[2]{\ColorBox{#1} #2}
\newcommand{\Heading}[1]{\textbf{#1.}}

\begin{document}

\title{\textls[-13]{ChangePrism:\! Visualizing the Essence of Code
Changes}}

\author{
\IEEEauthorblockN{%
  Lei Chen\IEEEauthorrefmark{1},
  Michele Lanza\IEEEauthorrefmark{2}, 
  Shinpei Hayashi\IEEEauthorrefmark{1}}
\IEEEauthorblockA{%
  \IEEEauthorrefmark{1}\emph{School of Computing, Institute of Science Tokyo, Japan} ~~
  \IEEEauthorrefmark{2}\emph{REVEAL @ Software Institute - USI, Lugano, Switzerland}}
}

\maketitle
\thispagestyle{plain}
\pagestyle{plain}
\begin{abstract}

Understanding the changes made by developers when they submit a pull request and/or perform a commit on a repository is a crucial activity in software maintenance and evolution. The common way to review changes relies on examining code diffs, where textual differences between two file versions are highlighted in red and green to indicate additions and deletions of lines. This can be cumbersome for developers, making it difficult to obtain a comprehensive overview of all changes in a commit. Moreover, certain types of code changes can be particularly significant and may warrant differentiation from standard modifications to enhance code comprehension.
We present a novel visualization approach supported by a tool named \textit{ChangePrism}, which provides a way to better understand code changes. The tool comprises two components: extraction, which retrieves code changes and relevant information from the git history, and visualization, which offers both general and detailed views of code changes in commits. The general view provides an overview of different types of code changes across commits, while the detailed view displays the exact changes in the source code for each commit.

\faIcon{video} Video demonstration: \url{https://youtu.be/jMoGLfM3KIM}
\end{abstract}


\section{Introduction}

Understanding code changes is a fundamental activity that underpins software maintenance and evolution. When developers submit pull requests or commit changes to a repository, it is essential to review and comprehend these modifications accurately. Traditionally, this review process relies on examining code diffs, where textual differences between two code snapshots are highlighted \cite{bacchelli2013expectations}. Additions and deletions are marked respectively in green and red, providing a basic visual cue to identify changes.

While widely used, relying solely on red and green highlights to indicate additions and deletions is insufficient for comprehensive code reviews. This simplistic approach fails to differentiate between various types of changes, such as newly introduced and removed code (additions and deletions), and changes to existing code (modifications). As a result, developers may overlook important nuances in the code, making it challenging to grasp the full scope and impact of changes. Furthermore, the textual diff representation can be cumbersome, especially when a single commit involves changes that are distributed across multiple files and therefore lacks a general view to help developers quickly grasp the scale of the changes. In addition, not all code changes are equal. Some modifications, such as refactorings or micro-changes \cite{chenlei-icsme2024}, carry different significance levels and impact on the overall system \cite{fowler2018refactoring}. The awareness of those modifications can help developers prioritize and understand the more critical aspects of a commit while reducing cognitive load \cite{rediffs,ge2017refactoring,alves2014refdistiller}.

We present \textit{ChangePrism}, a code diff visualization tool to enhance the understanding of code changes by automatically identifying various change types, including refactorings and micro-changes, and representing them with distinct visual elements. It comprises two main components: extraction and visualization. The extraction component retrieves code changes and relevant information from a Git history. The visualization component then presents the changes and information in multiple views: a general view showing the overview of changes of all commits, and a detailed view detailing the changes of each commit. By integrating semantic change information into the visual review process, ChangePrism enables developers to more effectively understand what changed in a commit, where those changes occur, and what type of changes they are.

\section{ChangePrism in a Nutshell}

\begin{figure*}[ht]\centering
    \includegraphics[width=0.75\linewidth]{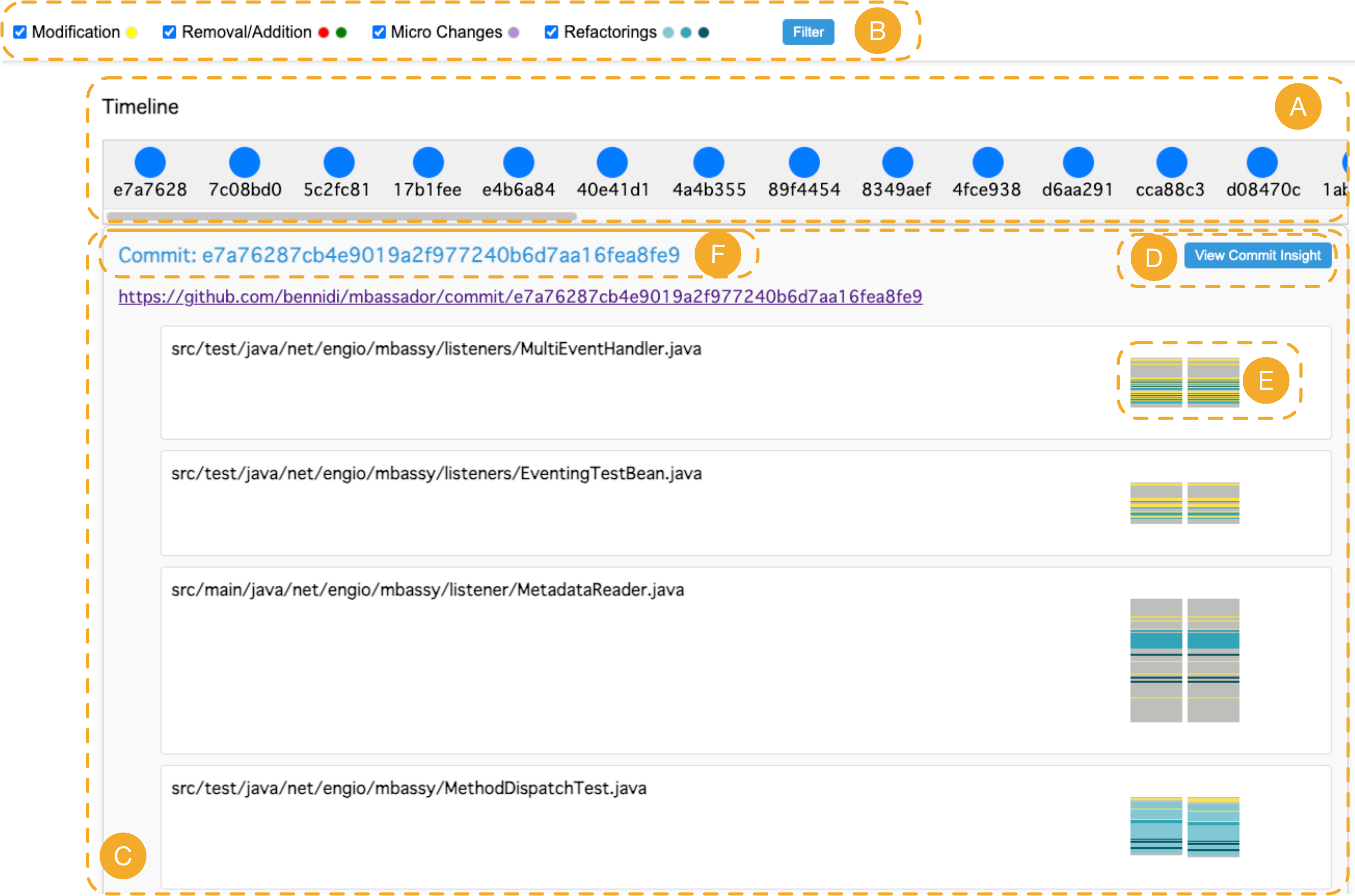}
    \caption{General view.}
    \label{f:generalview}
\end{figure*}

\begin{figure*}[ht]\centering
    \includegraphics[width=0.7\linewidth]{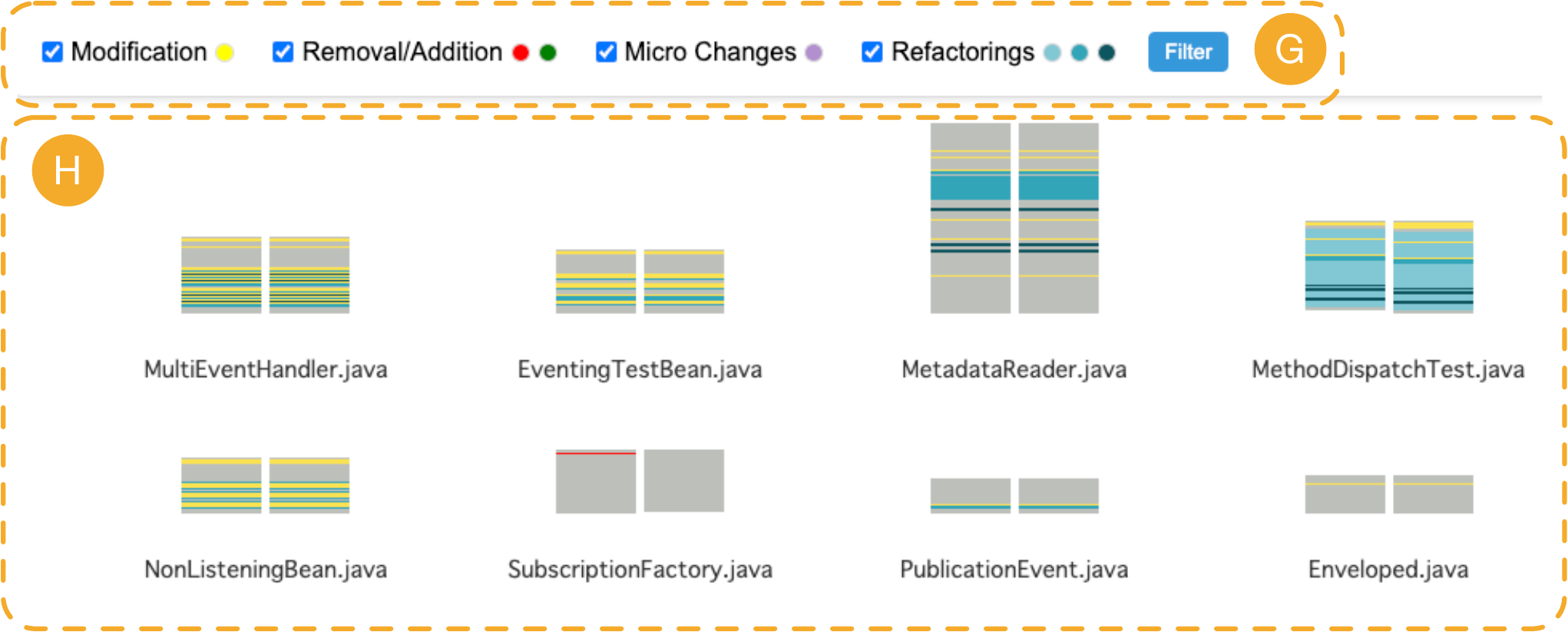}
    \caption{Commit Insight view.}
    \label{f:commitInsightView}
\end{figure*}

\begin{figure*}[ht]\centering
    \includegraphics[width=\linewidth]{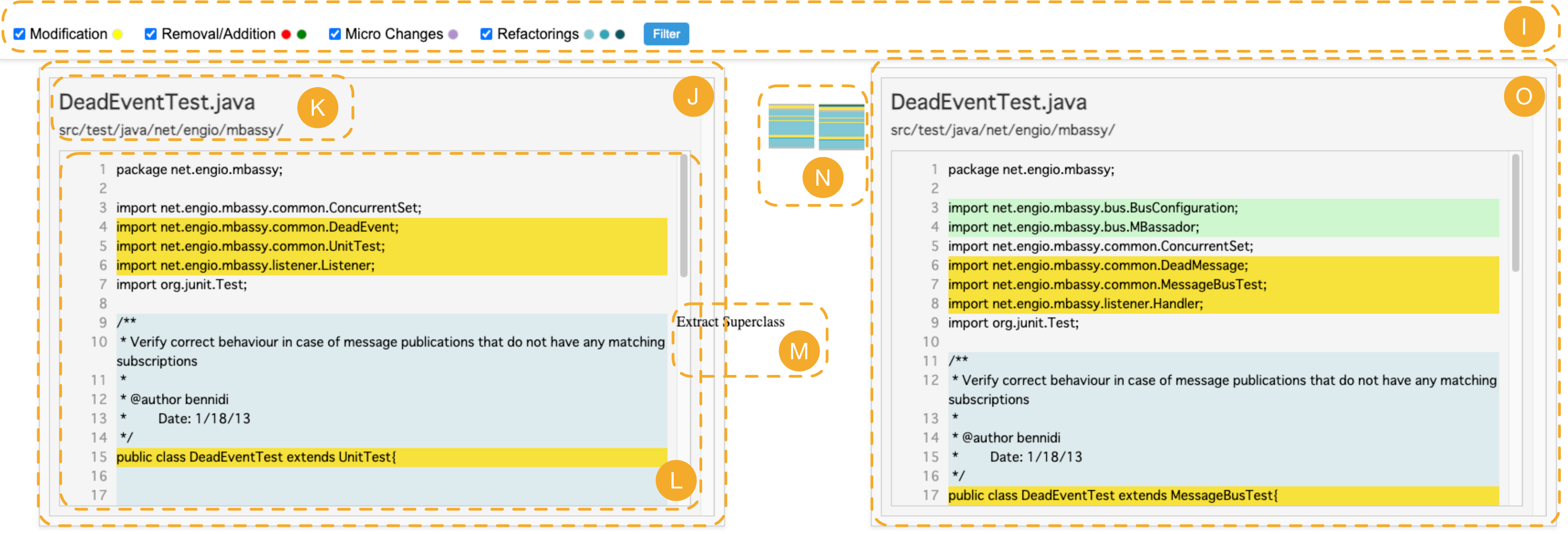}
    \caption{Code Detail view.}
    \label{f:codeDetailView}
\end{figure*}

ChangePrism offers three views: a) General view, offering a comprehensive summary of commits and changes, as well as entry points to the other two views; b) Commit Insight view, presenting a snapshot of the changes within a single commit; and c) Code Detail view, displaying specific changes with the corresponding source code. We present ChangePrism by utilizing it to review commits in the repository \textit{mbassador}.

\Heading{General View}
\Cref{f:generalview} shows the General view of ChangePrism. At the top is the timeline (\Circled{A}), which displays the commit history in the repository. Each commit is represented as a blue circle, with its SHA-1 hash shown below it. The commits are arranged chronologically from left to right. By scrolling the timeline horizontally, more commits can be revealed. Clicking on a commit will display its details on the page. Above the timeline, there are four checkboxes (\Circled{B}) that allow users to select the types of code changes they wish to view. This feature enables users to focus on specific types of changes.
The area \Circled{C} contains information about a single commit, including its SHA-1 hash of the commit, a link to the commit on GitHub, the paths of the changed files, and the \textit{ChangeSpectrum} for each file (\Circled{E}). The \textit{ChangeSpectrum} is a clear and concise representation of various types of code changes and is introduced in later. It provides users with a general view of changed files in commit and the scales of code changes. The button \Circled{D} is the entry point to access the \textit{Commit Insight View}, and the SHA-1 hash of the commit \Circled{F} is the entry point to access the \textit{Code Detail view} of that commit.

\Heading{Commit Insight View}
This view offers a visual summary of the \textit{ChangeSpectrum} within a single commit, as illustrated in \cref{f:commitInsightView}. The checkboxes (\Circled{G}) function in the same way as in the \textit{General view}, enabling users to filter the types of changes. The \Circled{H} depicts ChangeSpectrum in each changed file in the commit. This view ensures users have immediate access to the scope and impact of code changes in all files across the commit. Additionally, users are able to prioritize the review of the most vital or complex changed files, such as \textit{MultiEventHandler.java}, as suggested by Fregnan et al.~\cite{fregnan2022first}.

\Heading{Code Detail View}
This view (\cref{f:codeDetailView}) shows the concrete source code and the code changes for changed files in a commit.
Similar to the \textit{General View}, users can select the code change type they wish to see, as shown in \Circled{I}. The \Circled{J} and \Circled{O} denote the pre-commit and post-commit windows with a scroll bar to show the pre-commit and post-commit source code of a file, respectively. The file path is shown above as \Circled{K}. 
Within the source code window, different colors are used to highlight various types of code changes as shown in \cref{t:change_types}. Purely added code lines are highlighted in green, while purely removed code lines are highlighted in red. 

The modified code is highlighted in yellow. Additionally, two special types of code changes, refactorings and micro-changes, are highlighted in blue and purple, respectively. For refactorings, we use three shades of blue—from light to dark—to distinguish among class-level, method-level, and statement-level refactorings. Class-level refactoring refers to changes that apply to the entire class, such as \refactoring{Move Class} and \refactoring{Extract Superclass}. Similarly, method-level refactoring pertains to changes at the method level, and statement-level refactoring involves changes to fields or within the method.
When users hover the mouse over the blue (refactoring) or purple (micro-change) areas, the corresponding type names for the code changes in those areas are displayed. For instance, as shown in \Circled{M}, a tooltip appears indicating that the light blue highlighted code changes contain an \refactoring{Extract Superclass} refactoring.

\section{Technical Background}

\subsection{Data Extraction} \label{ss:data_extraction}

\begin{figure}[t]\centering
    \includegraphics[width=0.78\columnwidth]{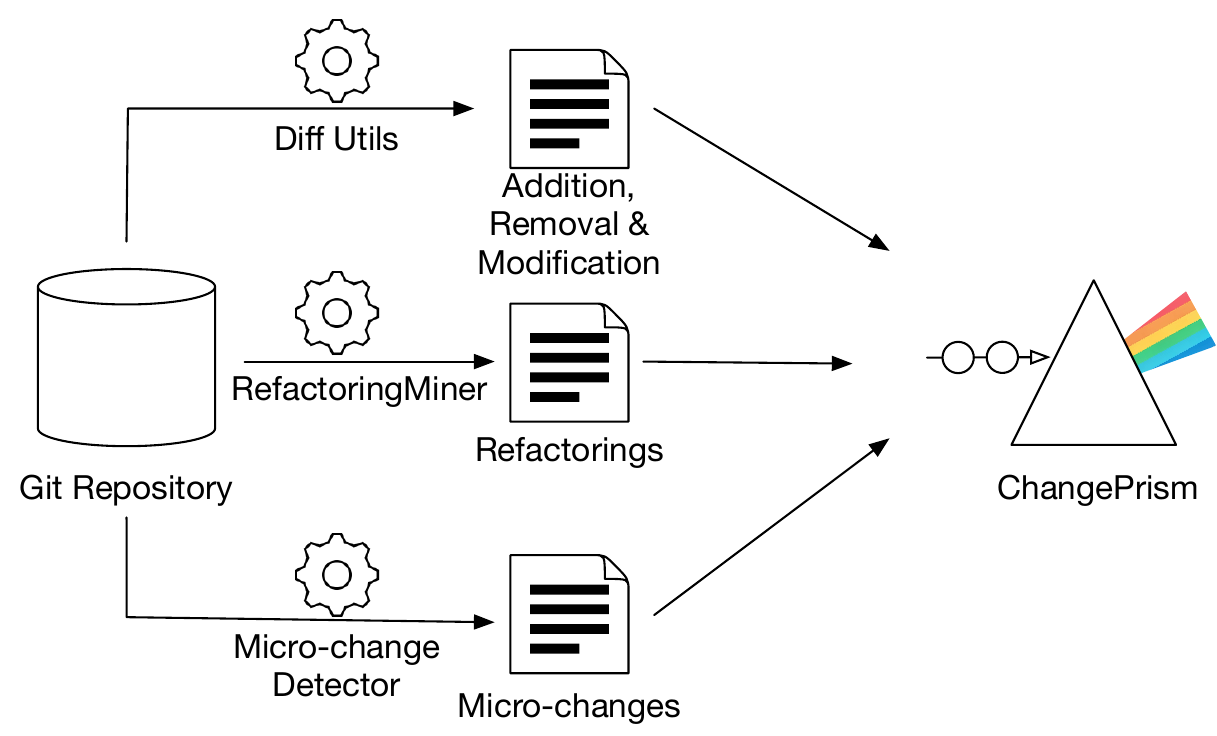}
    \caption{Data extraction.}\label{f:extraction}
\end{figure}

An overview of the data extraction process used for ChangePrism is shown in \cref{f:extraction}.
The data source is a Git repository, and the output consists of various types of code changes. The \textit{java-diff-utils} library\cite{java-diff-utils} is used to extract the addition, removal, and modification code changes. The state-of-the-art tool RefactoringMiner~\cite{Tsantalis:TSE:2020:RefactoringMiner2.0} is used to extract refactorings, and the micro-change detector~\cite{chenlei-icsme2024} is employed to extract micro-changes. The extracted result includes the change type names and their corresponding positions. They serve as the input to the ChangePrism.

\begin{figure*}[ht]\centering
    \includegraphics[width=0.95\linewidth]{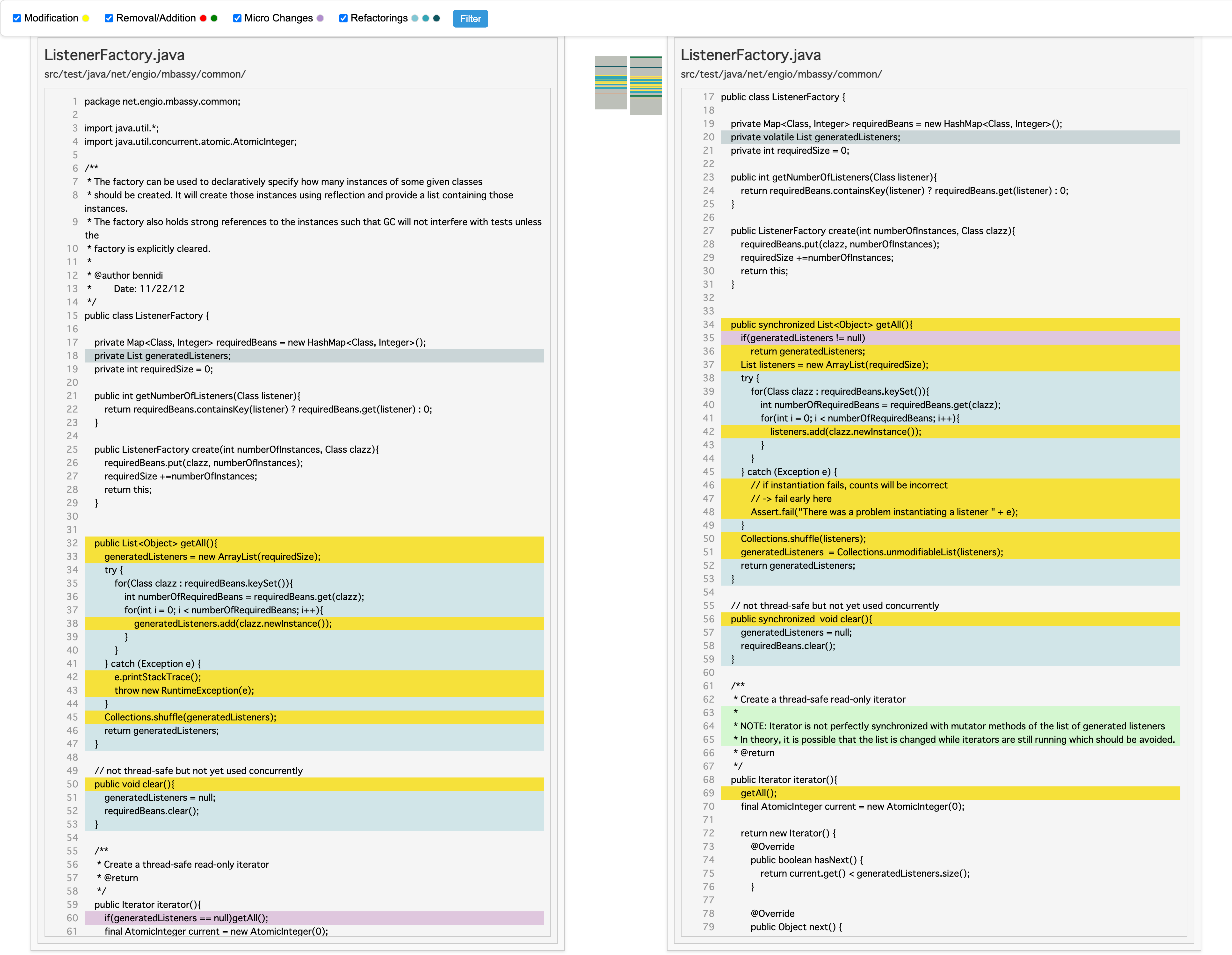}
    \caption{ChangePrism interface.}
    \label{f:casestudy}
\end{figure*}

\subsection{Change Spectrum} \label{ss:change_spectrum}

\textit{ChangeSpectrum} is a visualization tool designed to enhance the understanding of code changes between commits. It consists of two mini-maps representing the pre-commit and post-commit source code of a file. The height of each mini-map corresponds to the length of the file, providing a proportional view of the source code scale. The mini-map is constructed as a layered spectrum piled with different colors, each representing a specific type of code change. 

Different types are displayed in their relative positions on  ChangeSpectrum, allowing developers to see exactly where in the file the changes have occurred. This spatial representation aids in understanding the context of the changes within the overall file structure. In addition, changes are layered according to their priority. The priorities are displayed in \cref{t:change_types}.

\begin{table}[t]\centering
\caption{Supported Change Types}
\label{t:change_types}
\rowcolors{2}{gray!10}{white}
\begin{tabular}{cll}\hline
\textbf{Priority} & \textbf{Type} & \textbf{Color}\\\hline
1 & Class-level refactoring & \ShowColor{classRefactoringColor}{Light blue} \\
2 & Method-level refactoring & \ShowColor{methodRefactoringColor}{Blue} \\
3 & Modification & \ShowColor{modificationColor}{Yellow} \\
4 & Addition & \ShowColor{additionColor}{Green} \\
5 & Removal & \ShowColor{removalColor}{Red} \\
6 & Statement-level refactoring & \ShowColor{variableRefactoringColor}{Dark blue} \\
7 & Micro-change & \ShowColor{microChangeColor}{Purple} \\\hline
\end{tabular}
\end{table}

This prioritization helps developers quickly identify the most significant modifications in the code change.
\begin{itemize}
  \item \textit{Textual code diff} involves three types: addition, removal, and modification. The addition refers to introducing new lines of blocks, the removal involves deleting existing lines or blocks of code, and the modifications occur when an existing line or block of code is altered.
  \item \textit{Refactorings} refer to the code changes that enhance the internal structure of the code while preserving its external behavior~\cite{fowler2018refactoring},
 and it is a key practice in agile development processes~\cite{opdyke1992refactoring}.
  \item \textit{Micro-changes} refer to code change operations described in natural language, designed to bridge the cognitive divide by translating the textural diffs into more understandable natural-language described operations.
\end{itemize}

\section{Case Study: ChangePrism in Action} \label{s:casestudy}

We illustrate the use of ChangePrism to review a commit in the OSS~\textit{mbassador}\footnote{\url{https://github.com/bennidi/mbassador/commit/5c2fc81}}. There is only one file \textit{ListenerFactory.java} changed in this commit, which can be observed from both the \textit{General View} and the \textit{Commit Insight View}. The visualized code changes are shown in \cref{f:casestudy}. From the ChangeSpectrum in the middle of \cref{f:casestudy}, users can easily identify that the main changes occur in the middle of the file. The blue layers mixed with yellow layers indicate a method-level refactoring and additional changes to that method.

Additionally, the right mini-map shows a larger height with some green layers, indicating that new code has been added.

As shown in \cref{f:casestudy}, Lines 63--65 contain newly added Javadoc comments. These lines are highlighted in green, indicating that they belong to the addition category. Line 18 on the left side is highlighted blue, and it is a \refactoring{Add Attribute Modifier} refactoring, which adds the modifier keyword \sourcecode{volatile} to the class field \sourcecode{generatedListeners}. Lines 32--47 in the left side is the method \sourcecode{getAll()}. The method block is highlighted as light blue, and the tooltip indicates the existence of a method-level refactoring \refactoring{Add Method Modifier}. The method signature is highlighted in yellow on Lines 32 and 34 on the left and right windows, respectively, indicating the addition of the \sourcecode{synchronized} keyword, which is the refactoring \refactoring{Add Method Modifier}. Line 35 on the post-commit window is highlighted in purple color, and the tooltip says the existence of two micro-changes: \microchange{Insert Condition Block} and \microchange{Encapsulate In Condition}. 
The code block ranges in Lines 35--36 is the inserted condition block, and the class field \sourcecode{generatedListeners} in the is encapsulated into the conditional expression to check whether it is initialized in the commit.

In the subsequent yellow highlighted code, a local variable \sourcecode{listeners} is created to replace the \sourcecode{generatedListeners} in the pre-commit code, and the exception thrown is replaced by an assertion. With the different colors highlighting the code, the developer can effectively capture the main logic change.

Lines 50--53 in the pre-commit is the method \sourcecode{clear()}, and it is highlighted in light blue. From the tooltip, we can know this is a \refactoring{Add Method Modifier} refactoring, and from the method signature highlighted in yellow, we can know a keyword \sourcecode{synchronized} is added as a method modifier.

Line 60 in the pre-commit window is highlighted in purple with a tooltip indicating a micro-change~\microchange{Extract From Condition}. In the corresponding Line 69 in the post-commit window, in yellow, the method \sourcecode{getAll()} has been extracted from its previous position within the condition that checked whether \sourcecode{generatedListeners} was initialized. This method is now executed without any conditional checks.

\section{Related Work}

Visualization techniques have been extensively researched and developed to assist developers in understanding code changes. D'Ambros et al.\cite{d2010commit} proposed \textit{Commit 2.0}, which highlights the changes in packages, classes, and methods between the last version and the locally modified version before commit. Fregnan et al.\cite{fregnan2023graph} offer a summary of the contents in a pull request by presenting method calls, class relations, and code changes. Both studies provide visualizations at a coarse level, whereas ChangePrism offers a finer line-level visualization of changes. DiffViz~\cite{frick2018diffviz} proposed a tree-diff visualization at the line level but does not distinguish between different types of changes as ChangePrism does. Spike, proposed by Escobar et al.~\cite{escobar2022spike}, RAID proposed by Brito et al.~\cite{brito2021raid}, and RefactoringMiner\cite{Tsantalis:TSE:2020:RefactoringMiner2.0}, can highlight both refactorings and regular code changes. However, they do not provide an overview of changes in a commit, failing to give users a comprehensive grasp of changes across the commit as effectively as ChangePrism does. To our knowledge, no existing tool supports the visualization of micro-changes. ChangePrism is the first tool specifically designed to surface these fine-grained transformations within a commit.

Line-based change visualization tools have long used color highlights to represent code modifications across files and revisions. Seesoft\cite{seesoft} visualizes each line of code as a thin horizontal strip, using color to indicate statistical attributes such as change recency. CVSscan\cite{cvsscan} visualizes the evolution of each line across revisions along a time axis, effectively showing when each line was modified. CVSgrab\cite{cvsgrab} presents a timeline view of file-level activity, showing when and how frequently each file changed during the evolution. While these tools effectively reveal the location and frequency of line-level changes, they do not convey the type or semantic intent of the changes. Our approach builds upon these line-oriented visual encodings, but extends them by distinguishing a broader range of change types, including refactorings and micro-changes, enabling a richer understanding of code evolution.

\section{Conclusion}

We proposed a novel code visualization tool ChangePrism. We introduced its user interface and provided examples of how it works. Our future plans for ChangePrism are:

\Heading{Enhanced user interface} The user interface of ChangePrism can be made more user-friendly by adding several new features. A search function can be added to the code detail view, allowing users to search according to the type of change. The search results will be highlighted in the ChangeSpectrum. 
Additionally, a navigation feature can be implemented so that when a user clicks on a layer in the ChangeSpectrum, the pre-commit and post-commit windows automatically jump to the corresponding code.

\Heading{Improved accuracy in highlighting} We plan on enhancing the change type extraction process to achieve higher accuracy. For instance, currently, only the line where an attribute is renamed is highlighted for the \refactoring{Rename Attribute} refactoring. However, to help reviewers better understand the impact of this refactoring, the lines where the renamed attribute is used should also be highlighted.

\Heading{Support for additional types of code changes} The tool should support additional types of code changes, such as \textit{Behavioral Changes}, which alter the program's behavior, and \textit{Syntax Adjustments}, which include changes like adding a blank line or modifying a curly brace.

\Heading{Integration with GitHub and Review Platforms} Our tool should be integrated with GitHub and Review Platforms to enable a more effective review without altering the workflow reviewers and developers are already familiar with.

\section*{Acknowledgments}

This work is partly supported by the SNSF through the project ``FORCE'' (No. 232141),
the Young Researchers Exchange Programme under the Japanese-Swiss Science and Technology Programme, JST SPRING (JPMJSP2106), and JSPS KAKENHI (JP23K24823, JP25K03102, JP24H00692).  


\bibliographystyle{IEEEtran}
\bibliography{references}

\begin{thebibliography}{10}
\providecommand{\url}[1]{#1}
\csname url@samestyle\endcsname
\providecommand{\newblock}{\relax}
\providecommand{\bibinfo}[2]{#2}
\providecommand{\BIBentrySTDinterwordspacing}{\spaceskip=0pt\relax}
\providecommand{\BIBentryALTinterwordstretchfactor}{4}
\providecommand{\BIBentryALTinterwordspacing}{\spaceskip=\fontdimen2\font plus
\BIBentryALTinterwordstretchfactor\fontdimen3\font minus
  \fontdimen4\font\relax}
\providecommand{\BIBforeignlanguage}[2]{{%
\expandafter\ifx\csname l@#1\endcsname\relax
\typeout{** WARNING: IEEEtran.bst: No hyphenation pattern has been}%
\typeout{** loaded for the language `#1'. Using the pattern for}%
\typeout{** the default language instead.}%
\else
\language=\csname l@#1\endcsname
\fi
#2}}
\providecommand{\BIBdecl}{\relax}
\BIBdecl

\bibitem{bacchelli2013expectations}
A.~Bacchelli and C.~Bird, ``Expectations, outcomes, and challenges of modern
  code review,'' in \emph{Proc. ICSE}, 2013, pp. 712--721.

\bibitem{chenlei-icsme2024}
L.~Chen, M.~Lanza, and S.~Hayashi, ``Understanding code change with
  micro-changes,'' in \emph{Proc. ICSME}, 2024, pp. 363--374.

\bibitem{fowler2018refactoring}
M.~Fowler, \emph{Refactoring: Improving the Design of Existing Code}.\hskip 1em
  plus 0.5em minus 0.4em\relax Addison-Wesley Professional, 2018.

\bibitem{rediffs}
S.~Hayashi, S.~Thangthumachit, and M.~Saeki, ``{REdiffs}: Refactoring-aware
  difference viewer for {Java},'' in \emph{Proc. WCRE}, 2013, pp. 487--488.

\bibitem{ge2017refactoring}
X.~Ge, S.~Sarkar, J.~Witschey, and E.~Murphy-Hill, ``Refactoring-aware code
  review,'' in \emph{Proc. VL/HCC}, 2017, pp. 71--79.

\bibitem{alves2014refdistiller}
E.~L. Alves, M.~Song, and M.~Kim, ``{RefDistiller}: {A} refactoring aware code
  review tool for inspecting manual refactoring edits,'' in \emph{Proc. FSE},
  2014, pp. 751--754.

\bibitem{fregnan2022first}
E.~Fregnan, L.~Braz, M.~D'Ambros, G.~{\c{C}}al{\i}kl{\i}, and A.~Bacchelli,
  ``First come first served: {T}he impact of file position on code review,'' in
  \emph{Proc. ESEC/FSE}, 2022, pp. 483--494.

\bibitem{java-diff-utils}
``{java-diff-utils},''
  \url{https://java-diff-utils.github.io/java-diff-utils/}, 2017.

\bibitem{Tsantalis:TSE:2020:RefactoringMiner2.0}
N.~Tsantalis, A.~Ketkar, and D.~Dig, ``{RefactoringMiner} 2.0,'' \emph{IEEE
  Trans. Softw. Eng.}, vol.~48, no.~3, pp. 930--950, 2022.

\bibitem{opdyke1992refactoring}
W.~F. Opdyke, ``Refactoring object-oriented frameworks,'' Ph.D. dissertation,
  University of Illinois at Urbana-Champaign, 1992.

\bibitem{d2010commit}
M.~D'Ambros, M.~Lanza, and R.~Robbes, ``Commit 2.0,'' in \emph{Proc. 1st
  Workshop on Web 2.0 for Software Engineering}, 2010, pp. 14--19.

\bibitem{fregnan2023graph}
E.~Fregnan, J.~Fr{\"o}hlich, D.~Spadini, and A.~Bacchelli, ``Graph-based
  visualization of merge requests for code review,'' \emph{J. Syst. Softw.},
  vol. 195, pp. {111\,506}:1--20, 2023.

\bibitem{frick2018diffviz}
V.~Frick, C.~Wedenig, and M.~Pinzger, ``{DiffViz}: A diff algorithm independent
  visualization tool for edit scripts,'' in \emph{Proc. ICSME}, 2018, pp.
  705--709.

\bibitem{escobar2022spike}
R.~Escobar, J.~P.~S. Alcocer, H.~Tarner, F.~Beck, and A.~Bergel, ``{Spike} --
  {A} code editor plugin highlighting fine-grained changes,'' in \emph{Proc.
  VISSOFT}, 2022, pp. 167--171.

\bibitem{brito2021raid}
R.~Brito and M.~T. Valente, ``{RAID}: {T}ool support for refactoring-aware code
  reviews,'' in \emph{Proc. ICPC}, 2021, pp. 265--275.

\bibitem{seesoft}
S.~Eick, J.~Steffen, and E.~Sumner, ``{Seesoft}---a tool for visualizing line
  oriented software statistics,'' \emph{IEEE Trans. Softw. Eng.}, vol.~18,
  no.~11, pp. 957--968, 1992.

\bibitem{cvsscan}
L.~Voinea, A.~Telea, and J.~J. van Wijk, ``{CVSscan}: {V}isualization of code
  evolution,'' in \emph{Proc. SOFTVIS}, 2005, pp. 47--56.

\bibitem{cvsgrab}
S.~L. Voinea and A.~Telea, ``{CVSgrab}: Mining the history of large software
  projects,'' in \emph{Proc. EuroVis}, 2006, pp. 187--194.

\end{thebibliography}

\end{document}